\documentclass[aps]{revtex4} 

\usepackage{amsmath}
\usepackage{amssymb}
\usepackage[dvips]{graphicx}

\newcommand{\dd}  {\cos^2\sigma+e^{-4r_0}\sin^2\sigma}
\newcommand{\ee}  {e^{-2r_0}}

\begin{document}
{\small The original publication is available at www.springerlink.com }   
 \vspace{ 1cm }
\title{Stability of Closed Timelike Curves in the G\"odel Universe}

 \author{Val\'eria M. Rosa\footnote{e-mail: vmrosa@ufv.br}
  }

 \affiliation{
Departamento de Matem\'atica, Universidade Federal de Vi\c{c}osa, 
36570-000 Vi\c{c}osa, M.G., Brazil
}

\author{ Patricio  S. Letelier\footnote{e-mail: letelier@ime.unicamp.br}
 } 
 
\affiliation{
 Departamento de Matem\'atica Aplicada-IMECC,
Universidade Estadual de Campinas,
13081-970 Campinas,  S.P., Brazil}

\begin{abstract}
  We study, in some detail, the linear stability of closed timelike
  curves in the G\"odel universe. We show that these curves are stable.
  We present a simple extension (deformation) of the G\"odel metric
  that contains a class of closed timelike curves similar to the ones
  associated to the original metric. This extension correspond
  to the addition of matter whose energy-momentum tensor is analyzed.
  We find the conditions to have matter that satisfies the usual
  energy conditions. We study the stability of closed timelike curves
  in the presence of usual matter as well as in the presence of exotic
  matter (matter that does satisfy the above mentioned conditions). We
  find that the closed timelike curves in the G\"odel universe with 
 or without the inclusion of regular or  exotic matter are stable 
under  linear perturbations. We also find  a sort of structural stability.

\vspace{0.5cm}

\end{abstract}

\maketitle

\section{Introduction}

The G\"odel universe, a spacetime in which the matter takes the form
of a pressure-free perfect fluid with a negative cosmological
constant, is the most celebrated solution of Einstein field equations
that contains closed timelike curves (CTCs). This spacetime has a
five-dimensional group of isometries which is transitive. The matter
everywhere rotates relative to the compass of inertia with the angular velocity proportional to the square root of the matter density~\cite{goedel}~\cite{hawking}.
Dynamical conditions for time traveling in this spacetime  are not sufficient to exclude the existence of CTCs~\cite{pfarr}.

The G\"odel metric has some qualitative features  like the projections
of geodesics onto the 2-surface $(r,\phi)$ being simple closed
curves. This property can be  extended to a set of metrics called
G\"odel-type. It is possible to show explicitly that when the
characteristic vector field that defines a G\"odel-type metric is also
a Killing vector  we have closed timelike or null curves~\cite{gleiser}.
The study of the geodesic motion of free test particles in these
G\"odel-type spacetimes can be   extended to a family of
 homogeneous G\"odel-type spacetimes~\cite{calvao}.  
 G\"urses et al.~\cite{gurses} introduce and used
G\"odel-types metrics to find charged dust solutions to the Einstein
field equations in D dimensions.
  In the G\"odel spacetime timelike geodesics are not closed  as was
 independently proved by Kunt~\cite{kundt} and Chandrashekhar and
 Wrigth~\cite{chandra}. However there exist  timelike curves subject to
 an external force that are closed.

Exact solutions of Einstein-Maxwell equations that contain CTCs at
least for some values of the parameters are studied in ~\cite{klepac}.
It turns out that magnetic fields can give rise to non-trivial chronology violations.  A sufficiently large magnetic field can always ensure chronology violation.
The spacetime of an infinite rotating cylindrical shell of charged
 perfect fluid contains CTCs~\cite{gersl}. For the  conditions for the existence of  CTCs in the spacetime associated to a rigidly rotating cylinder of charged dust see
 Ivanov~\cite{ivanov}.  In spacetimes with conic singularities that
 represent cosmic strings we can have CTCs for one spinning
 string~\cite{ozdemir} and  for two parallel moving strings~\cite{gott}.

The stability of G\"odel's cosmological model with respect
 to scalar, vector, and
tensor perturbation modes using a gauge covariant formalism is
considered in~\cite{barrow}. It is found that the background vortical
energy contributes to the  gravitational pull of matter, while gradients
add to the pressure support. The balance between these two agents
effectively determines the stability of G\"odel universe against
matter aggregations.

The possibility to have  no violation of causality for  geodesics that in  finite intervals of time  goes back in time ($dt/ds<0$) was considered by~\cite{novello}. 
As it was already mentioned, the existence of CTCs contradicts the usual notion of causality. Beyond the usual paradoxes, it seems to induce physical
impossibilities, like the necessity to work with negative energy
densities. One could speculate that these impossibilities will be
eliminated by quantum-gravitational effects.  All our experience 
seems to indicate that the physical laws do not allow  the
appearance of CTCs. This is that, essentially,  says the Chronology
Protection Conjecture (CPC) proposed by Hawking in
1992~\cite{hawking2}.

Other possible explanation for CTCs is that the metrics studied until
now are unrealistic~\cite{bonnor2}.  But, there exists an exact,
asymptotically flat solution of vacuum Einstein equations which
contains CTCs that can represent the exterior of a spinning rod of
finite length~\cite{bonnor}.

The existence of CTCs, in principle, should be a matter of experimentation. If 
the General Relativity predicts them in a physically reasonable situation and they
are not found, we will have that  this theory is in trouble. If they are found we will have a bigger problem, our usual notion of causality will need a deep revision.

The possibility that a spacetime
associated to a realistic model of matter may contain CTCs leads us to
ask how permanent is the existence of these curves. Perhaps, one may
rule out the CTCs by simple considerations about their linear
stability. Otherwise, if these curves are stable under linear
perturbations the conceptual problem associated  to their existence is enhanced.

In this first paper about linear stability of CTCs we consider the
G\"odel case due to its historical relevance (first paradigm for
CTCs) as well as its mathematical simplicity. We intend to study, in
subsequents works, the stability of CTCs in G\"odel-type
metrics~\cite{gurses, gleiser}, Bonnor metric ~\cite{bonnor} and other
relevant spacetimes. All these metrics presents their own
peculiarities that deserve a special treatment.

 The paper is divided as follows, in Section 2, we review in some
 detail the CTCs in the G\"odel metric.  In particular we present the CTCs
 and its corresponding forces in the usual G\"odel coordinates as
 well as in Cartesian coordinates.  In Section 3 we study the
 stability of CTCs when they suffer a small perturbation. We solve the
 linear system of equations for the evolution of the perturbation in
 the original G\"odel coordinates in an exact form. For future
 reference the same system in Cartesian coordinates is solved
 numerically for a representative range of initial conditions. In
 Section 4 we present a simple extension (deformation) of the G\"odel
 metric that corresponds to the addition of matter. We look for the
 conditions for geodesics to be timelike and to be closed. We prove that
 these two conditions can not be satisfied at the same time. We present
 the CTCs in an explicit form together with their associated forces.
We find that the addition of  matter either usual or exotic does not change 
 the stability of the G\"odel CTCs. In the last section we
 summarize and discuss some of our results.

\section{G\"odel metric, geodesics and CTCs}
In this section we review the CTCs in G\"odel metric following mainly
references~\cite{goedel}~\cite{pfarr}~\cite{kundt}. 
In standard coordinates, $\bar{X}^{\mu}=[\bar{t},r\maketitle,\varphi,z]$,
the G\"odel metric is~\cite{pfarr},
\begin{equation}
  ds^2 = \frac{4}{\beta^2} \left[ 
    d\bar{t}^2-dr^2
    +\big( \sinh^4r-\sinh^2r \big) \;d\varphi^2
    +2\sqrt{2}\sinh^2 r \;dt \,d\varphi \;\right]
  -dz^2,
  \label{metricaGoedel}
\end{equation}
where $\beta$ is a constant that describes the vorticity of the 
four-vector $u^{\mu} =[1,0,0,0] $, $-\infty < \bar{t} < \infty$,
$-\infty < z < \infty$,
$r>0$ and $\varphi \in [-\pi,\pi]$. The limits $\varphi=-\pi$ and
 $\varphi=\pi$ are topologically identified.

The metric~(\ref{metricaGoedel}) satisfies  the Einstein  field equations, 
\begin{equation}
  R_{\mu\nu}
  - \frac{1}{2} R \,g_{\mu\nu} +\Lambda\,g_{\mu\nu}
  = -T_{\mu\nu},
\end{equation}
with $T^{\mu\nu}=\rho \,u^{\mu}u^{\nu}$, $\Lambda=-\frac{\beta^2}{2}$ and
$\rho=\beta^2$. 
We use units such that $c=8\pi G=1$.

Let  us denote by $\gamma$  the  particular closed curve  given in its parametric form by, 
\begin{equation}
  \bar{t} =   \bar{t}_0, \hspace{2cm}
  r       =   r_0,       \hspace{2cm}
  \varphi    \in [-\pi,\pi], \hspace{2cm} z=0,
  \label{CTCGoedel}
\end{equation}
where $t_0$ and $r_0$ are constants.  When $\gamma$ is parametrized with an arbitrary 
variable $u$ the condition to be timelike is 
$\frac{d\bar{X}^{\mu}}{du}\frac{d\bar{X}_{\mu}}{ du}>0 $. The curve  $\gamma$  is timelike when $r_0 >\log(\sqrt{2}+1)$.  The  four-acceleration of this CTC is
\begin{equation}
  \label{forceCTCPfarr}
  \bar{a}^\mu = \delta^\mu_1 \;\sinh r_0 \;\cosh r_0 \;
          \big( 2\cosh^2 r_0 - 3 \big) \;\dot{\varphi}^2.
\end{equation}
In the ``Cartesian''   coordinates $X^{\mu}=(t,x,y,z)$   defined by
\begin{eqnarray}
&&  y  = \big( \cosh(2r) + \sinh(2r) \cos\varphi \big)^{-1}, \\
&& \beta x =  \big( \sinh(2r)   \;        \sin\varphi \big) \;y,  \\
&&  \frac{\varphi}{2}+(\,\frac{\beta t-2\bar{t}}{2\sqrt{2}}\,) 
     =  \arctan \left( \,e^{-2r}\tan\frac{\varphi}{2} \;\right),\\
&&z=z,
  \label{transformacaoCoordenadas}
\end{eqnarray}
with $-\infty < t < \infty$, $-\infty < x < \infty$, $-\infty <
 z < \infty$, and $\infty>y>0$ 
we find that the line element~(\ref{metricaGoedel}) reduces to
\begin{equation}
  ds^2 = \left[ dt + \frac{\sqrt{2}\; dx}{\beta y} \right]^2
       - \frac{ dx^2 + dy^2 } { (\beta y)^2 }
       - dz^2.
  \label{metricaPfarr}
\end{equation}
 In these coordinates the geodesic equation gives us,
\begin{eqnarray}
&&\ddot{t} = \frac{2}{y}\;\dot{t}\dot{y}+ \frac{\sqrt{2}}{\beta
y^2}\;\dot{x}\dot{y},\\ &&\ddot{x} = -\beta \sqrt{2} \;\dot{t}\dot{y},\\
&&\ddot{y} = \beta \sqrt{2} \;\dot{t}\dot{x} +
\frac{\dot{x}^2+\dot{y}^2}{y},\\ &&\ddot{z} = 0,
  \label{SistemaPfarr}
\end{eqnarray}
where the overdot indicates derivation with respect to the proper time $s$.
This  system of equations admits the first integrals,
\begin{eqnarray}
  \label{first_integral_t}
&&\dot{t} = \frac{2y-y_0}{\sqrt{2}C},\\
  \label{first_integral_x}
&&\dot{x} = \frac{\beta y}{C}(y_0-y),\\
  \label{first_integral_y}
&&\dot{y} = \frac{\beta y}{C}(x-x_0),\\
  \label{first_integral_z}
&&\dot{z} = \frac{d}{C},
\end{eqnarray}
where $x_0$, $y_0$, $d$ and $C$ are integration constants. 
The solution of this system is
\begin{eqnarray}
  \label{solucaoGeodesicaG3_t}
&&t = \frac{\sqrt{2}}{\beta}
      \left[ 2 \arctan\left( \sqrt{N}\tan\sigma \right)
             - \frac{N+1}{2\sqrt{N}}\,\sigma \right]
    + t_0,\\
  \label{solucaoGeodesicaG3_x}
&&x = \frac{ 2 \sqrt{N}\,\eta\sin\sigma \,\cos\sigma}
           { \cos^2\sigma+ N \sin^2\sigma } + x_0,\\
  \label{solucaoGeodesicaG3_y}
&&y = \frac{y_0-\eta}
          {\cos^2\sigma+N\sin^2\sigma},\;\\
  \label{solucaoGeodesicaG3_z}
&&z = \frac{2\sigma \  d}{\beta \sqrt{y_0^2-\eta^2}}+z_0,
\end{eqnarray}
with 
\begin{eqnarray}
  \label{def_sigma}
  &&\sigma=\frac{\beta\sqrt{y_0^2-\eta^2}}{2C}(s-s_0), \\
  \label{def_N}
 && N=\frac{y_0-\eta}{y_0+\eta},
\end{eqnarray}
where $t_0$ is an integration constant, and $\eta$ is given by 
 $\eta^2:=(x-x_0)^2+(y-y_0)^2$.

The relation  $\dot{X}^{\mu}\dot{X}_{\mu}=1$ (with  $z=0$) reduces to 
\begin{equation}
  \label{timelike_condition}
  \left( \dot{t} + \frac{\sqrt{2}}{\beta y}\dot{x} \right)^2
 -\frac{ \dot{x}^2 + \dot{y}^2}{(\beta y)^2} 
 = 1.
\end{equation}
From~(\ref{first_integral_t})-(\ref{first_integral_y}) we get
\begin{equation}
  y_0^2 = 2 \big( C^2 + \eta^2 \big).
\end{equation}
Therefore
\begin{equation}
  \label{closed_condition}
  y_0^2 > 2\eta^2.
\end{equation}

The condition for a geodesic to be closed, i.e.,  
$t(-\frac{\pi}{2})=t(\frac{\pi}{2})$,
$x(-\frac{\pi}{2})=x(\frac{\pi}{2})$ e
$y(-\frac{\pi}{2})=y(\frac{\pi}{2})$, is
\begin{equation}
4\sqrt{N}=N+1\  \Rightarrow \  y_0^2=\frac{4}{3}\eta^2
\end{equation}
that contradicts (\ref{closed_condition}). Hence we have no timelike closed
geodesics in the G\"odel universe.

In order to obtain a CTC  in  Cartesian
coordinates  we use another differential equation for $t$ obtained
from (\ref{timelike_condition}) and equations
(\ref{solucaoGeodesicaG3_x}) and (\ref{solucaoGeodesicaG3_y}). Then
\begin{equation}
  \left[ \dot{t}+\frac{\sqrt{2}}{C}(y_0-y) \right]^2
  =\frac{C^2+\eta^2}{C^2},
  \label{conditionG3CTCtt}
\end{equation}
whose solution is
\begin{equation}
  t = \frac{2\sqrt{2}}{\beta} 
  \left[
    \arctan \left(\sqrt{N} \tan\sigma \right)
    -\frac{y_0\sqrt{2}-\sqrt{C^2+\eta^2}}
          {\sqrt{2(y_0^2-\eta^2)}}
     \sigma \right] + t_0.
  \label{solucaoCTCPfarr}
\end{equation}
These timelike curves are  closed when
\begin{equation}
C^2 = 2(y_0-\sqrt{y^2_0-\eta^2})^2 - \eta^2.
\label{closedCTCcondition}
\end{equation}
By replacing these parametric equations for the CTC into the geodesic
equations we find that the four-acceleration satisfies the relations
\begin{equation}
  a^0 = -(x-x_0)\frac{\lambda \sqrt{2}}{\beta}, \;\;\;
  a^1 =  \lambda y (x-x_0),                     \;\;\;
  a^2 = -\lambda y (y_0-y),                     \;\;\;
  a^3=0,
\label{forceCTC}
\end{equation}
where
\begin{equation}
 \lambda =\frac{\beta^2}{C^2} 
\left[ \sqrt{2(C^2+\eta^2)} - y_0 \right].\end{equation}
The right hand side
of~(\ref{forceCTC}) can be interpreted as the components of 
a specific external force
$F^\mu$ associated to  $\gamma$.

It is instructive to compare the force given in these two systems of
 coordinates.  In G\"odel standard coordinates we have
 $g_{\varphi\varphi}\dot{\varphi}^2=1$. Therefore  along the CTC $\gamma$
we have
\begin{equation}
 \dot{\varphi}^2 = \frac{\beta^2}{4 \big( \sinh^4r_0-\sinh^2r_0
 \big)}.\end{equation}
From equation~(\ref{forceCTCPfarr}) we can write the non zero component
of the force in following way,
\begin{equation}\bar{F}^1 = \frac{\beta^2 \sinh r_0 \cosh r_0 
\big( 2\cosh^2r_0 - 3 \big) }{ 4 \big( \sinh^4r_0-\sinh^2r_0 \big) } =
                   \frac{\beta^2 \sinh 2r_0 \big( \cosh 2r_0 - 2 \big)
                   } { 2 \big( (\cosh 2r_0-2)^2 - 1\big)
                   }.\end{equation}
The non zero components of force in Cartesian coordinates are,
\begin{eqnarray}
&&F^0 = -\frac{2\sqrt{2}}{\eta\beta}(x-x_0)\bar{F}^1,  \\
&&F^1 = \frac{2y}{\eta} (x-x_0)  \bar{F}^1, \\
&&F^2 = -\frac{2y}{\eta} (y_0-y) \bar{F}^1,   \\
&&F^3 = 0.
\label{forcaTransformada}
\end{eqnarray}

On  $\gamma$  the constants in~(\ref{def_sigma}) and
(\ref{def_N}) are now  $N=e^{-2r_0}$, $y_0=\cosh 2r_0$ and $\eta=\sinh
2r_0$. Therefore
\begin{equation}
\bar{F}^1 =  \frac{\beta^2 \eta  \big( y_0 - 2 \big) }
                   { 2 \big( (y_0-2)^2 - 1\big) }.\end{equation}
By using the condition~(\ref{closedCTCcondition}) we can cast $\lambda$ as
\begin{equation}
\lambda=\frac{\beta^2(y_0-2)}{(y_0-2)^2-1}.
\end{equation} 
Hence
  \begin{equation}
 \bar{F}^1 = \frac{\eta}{2}\lambda, \end{equation}
that is  force calculated in (\ref{forceCTC}).

\section{Linear perturbation of CTCs in the G\"odel universe}

A generic CTC $\gamma$ satisfies the system of equations given by 
\begin{equation}
\frac{D}{ds} \dot{X}^{\mu}= F^\mu(X),
\label{CTCsystem}
\end{equation}
where $\frac{D\  ()^{\alpha}}{ds} $ is the covariant derivative of
the vector field $()^{\alpha}$ along $\gamma(s)$ and $F^\mu$ is a
 given external  force.

Let $\tilde{\gamma}$ be the curve obtained from $\gamma$ after a
perturbation ${\bf \xi}$,
i.e.,$\tilde{X}^{\mu}=X^{\mu}+\xi^{\mu}$. Let ${e_{\alpha}}$ be a
given basis. In this basis ~(\ref{CTCsystem}) is represented
by the equation $\frac{D{\bf u}}{ds}={\bf F}$, where $\frac{D{\bf
u}}{ds}=(\frac{du^{\alpha}}{ds}+\Gamma^{\alpha}_{\beta
\mu}u^{\beta}u^{\mu})e_{\alpha}$, ${\bf F}=F^{\alpha}e_{\alpha}$,
${\bf u}=u^{\alpha}e_{\alpha}$ and $u^{\alpha}=\dot{X}^{\alpha}$.
In order to find the behavior of  ${\bf \xi}$ we
calculate the variation, in first approximation,  of both sides of
 $\frac{D{\bf u}}{ds}={\bf F}$. We find,  
\begin{eqnarray}
&&\frac{\delta D{\bf u}}{d s}=
(\frac{d^2\xi^{\alpha}}{ds}+2\Gamma^{\alpha}_{\beta \mu}\delta
u^{\beta}u^{\mu}+\Gamma^{\alpha}_{\beta
\mu,\lambda}\xi^{\lambda}u^{\beta}u^{\mu})e_{\alpha}+F^{\alpha}\delta
e_{\alpha}\\ &&\delta{\bf
F}=F^{\alpha}_{,\beta}\xi^{\beta}e_{\alpha}+F^{\alpha}\delta
e_{\alpha}.
\end{eqnarray}
where $()_{,\lambda}=\frac{\partial ()}{\partial x^\lambda}$.
Comparing the last two equations we get the system of differential equation
satisfied by the perturbation ${\bf \xi}$ 
\begin{equation}
\frac{d^2\xi^{\alpha}}{ds}+2\Gamma^{\alpha}_{\beta \mu}\delta
u^{\beta}u^{\mu}+\Gamma^{\alpha}_{\beta
\mu,\lambda}\xi^{\lambda}u^{\beta}u^{\mu}=F^{\alpha}_{,\beta}\xi^{\beta}.
\label{systemPerturbation}
\end{equation}
This last equation can be cast in a manifestly covariant form by 
noticing that its left-hand-side  is the well known geodesic
 deviation equation as pointed out in ~\cite{shirokov} and 
its right-hand-side is the Lie derivative of the force along $\xi^\mu$.

In standard coordinates $\bar{X}^{\mu}=[\bar{t},r,\varphi,z]$, the
system~(\ref{systemPerturbation}) reduces to
\begin{equation}
\begin{array}{l}
\ddot{\xi}^0+a\,\dot{\xi}^1=0,\\
\ddot{\xi}^1+b\,\dot{\xi}^0 +c\,\dot{\xi}^2+d\,\xi^1=0,\\
\ddot{\xi}^2+e\,\dot{\xi}^1=0,\\ \ddot{\xi}^3=0,
\end{array}
\label{sistemaCTCGoedel}
\end{equation}
where 

\begin{eqnarray}
a&=&(\beta\,\sinh^2r_0\,\sqrt{2})/(\cosh\,r_0\sqrt{\sinh^2r_0-1}), \nonumber\\
b&=&\beta\sqrt{2}\cosh\,r_0/\sqrt{\sinh^2r_0-1},\nonumber\\
 c&=&\beta\cosh\,r_0(2\cosh^2r_0-3)/\sqrt{\sinh^2r_0-1}, \nonumber\\ 
d&=&\beta^2\cosh^2r_0(2\cosh^2r_0-3)^2/(2\sinh^2r_0(\sinh^2r_0-1)^2),\nonumber\\
 e&=&\beta/(\sinh r_0\, \cosh r_0\sqrt{\sinh^2r_0-1}) \label{abc1} 
\end{eqnarray}

The solution of system of equations (\ref{sistemaCTCGoedel}) is,
\begin{equation}
\begin{array}{l}
\xi^0=-a(c_3\sin(\omega s+c_4)/\omega+\tau s)+c_1\,s+c_5,\\ 
\xi^1=c_3\cos(\omega s+c_4)+\tau, \\
\xi^2=-e(c_3\sin(\omega s+c_4)/\omega+\tau s)+c_2\,s+c_6,\\ 
\xi^3=c_7s+c_8,
\end{array}\label{exactpert}
\end{equation}
where $c_i,\;i=1,\dots,8$ are integration constants, $\omega=\sqrt{d-ab-ce}$,  and
$\tau = -(bc_1+cc_2)/\omega^2$.  In order that the perturbed curve, $\tilde\gamma$, remains on the plane $z=0$ 
we take initial conditions such that $c_7=c_8=0$, i.e., $\xi^3=0$.
The solution  (\ref{exactpert}) shows  the typical behavior  for    stability,
we have vibrational modes untangled with translational ones that can be 
eliminated by a suitable choice of the initial conditions.

For  future reference, we study the linear stability of the CTCs
in Cartesian coordinates. In these coordinates the
system~(\ref{systemPerturbation}) can be written as
\begin{eqnarray}
& &\ddot{\xi^0} =  a_{00} \dot{\xi}^0
               + a_{01} \dot{\xi}^1
               + a_{02} \dot{\xi}^2
               + b_{01} \xi^1 
               + b_{02} \xi^2  \nonumber \\
& &\ddot{\xi^1} =  a_{10} \dot{\xi}^0
               + a_{11} \dot{\xi}^1
               + a_{12} \dot{\xi}^2
               + b_{11} \xi^1 
               + b_{12} \xi^2  \nonumber \\
& &\ddot{\xi^2}  =  a_{20} \dot{\xi}^0
               + a_{21} \dot{\xi}^1
               + a_{22} \dot{\xi}^2
               + b_{21} \xi^1 
               + b_{22} \xi^2  \nonumber \\
& &\ddot{\xi^3}  = 0,
\label{systemPerturbation2}
\end{eqnarray}
where $(a_{ij})$ and $(b_{ij})$ are given by
\begin{equation}
(a_{ij}) = \frac{1}{C}\left[ \begin{array}{ccc}
\displaystyle  \frac{2 \beta S \ee}{D}                         &
\displaystyle  \sqrt{2}\,S                                     & 
\displaystyle  \frac{\sqrt{2}}{2}\big(1+(y_0-2)De^{2r_0}\big)  \\[5mm]
\displaystyle -\frac{\sqrt{2}\,\beta^2 S e^{-4r_0}}{D^2}       & 
0                                                              & 
\displaystyle -\beta\left( \frac{\ee}{D}-1 \right)             \\[5mm]
\displaystyle  \frac{\sqrt{2}\,\beta^2 \ee}{2D}
               \left(y_0-\frac{e^{-2r_0}}{D} \right)           & 
\displaystyle  \beta(y_0-1)                                    &
\displaystyle  \frac{2 \beta S \ee}{D}
\end{array}\right],
\end{equation}

\begin{equation}
(b_{ij}) = \left[ \begin{array}{cc}
\displaystyle  - \frac{\lambda\sqrt{2}}{\beta}                 &
\displaystyle  - \frac{y_0-1}{C^2}4\sqrt{2}S                   \\[5mm]
\displaystyle  \frac{\ee\;\lambda}{\dd}                        & 
\displaystyle  \frac{2\ee\;S\;\lambda}{\dd}                    \\[5mm]
\displaystyle  0                                               & 
\displaystyle  \lambda \left[ \frac{2\ee}{\dd}-y_0 \right]-
\frac{\eta^2\beta^2}{C^2}
\end{array}\right],
\end{equation}
and
\begin{equation}  D = \dd, \;\;\;
 S = \eta \sin\sigma \cos\sigma.
 \end{equation}
 As before, we take $\xi^3=0$ in order that $\tilde\gamma$ does not leave the plane $z=0$.
 
To describe the behavior of perturbed curve $\tilde{\gamma}$
we introduce two distance and one angle functions,
\begin{eqnarray}
R_2^2&=&(x-x_0)^2+(y-y_0)^2,\label{circle}\\
R_3^2&=&(t-t_0)^2+(x-x_0)^2+(y-y_0)^2,\label{circle3D}\\
\phi&=&\arctan\frac{y-y_0}{x-x_0}. \label{phi}
\end{eqnarray}
The first function  is a constant equal to $\eta$ when $x$ and $y$ are
 on the CTC $\gamma$. The second represent  a  ``radius" in spacetime and 
the third one is an angle on the usual space.
The variation of these functions along $\gamma(s)$ are,
\begin{eqnarray}
\delta R_2&=&[(x-x_0)\xi^1+(y-y_0)\xi^2]/\eta, \label{deltaR2}\\
\delta R_3 &=&[(t-t_0)\xi^0+ (x-x_0)\xi^1+(y-y_0)\xi^2]/R_3, \label{deltaR3}\\
\delta \phi&=&[(x-x_0)\xi^2-(y-y_0)\xi^1]/\eta.\label{deltaPhi}
\end{eqnarray}

To study these functions we solve the system  (\ref{systemPerturbation2})
by running the independent variable  $\sigma$, first from $-\pi/2$ to $0$ and second from $\pi/2$ to $0$, we recall  that the points $-\pi/2$  and $\pi/2$  on the curve
 are identified. We keep
the initial position of the perturbation equal to zero and analyze the
behavior of $\delta R_2,\delta \phi$ and $\delta R_3$. The perturbation  initial velocity is taken each time with only one component different from zero.

\begin{figure}[!ht]
 \centering
 \includegraphics[scale=.5]{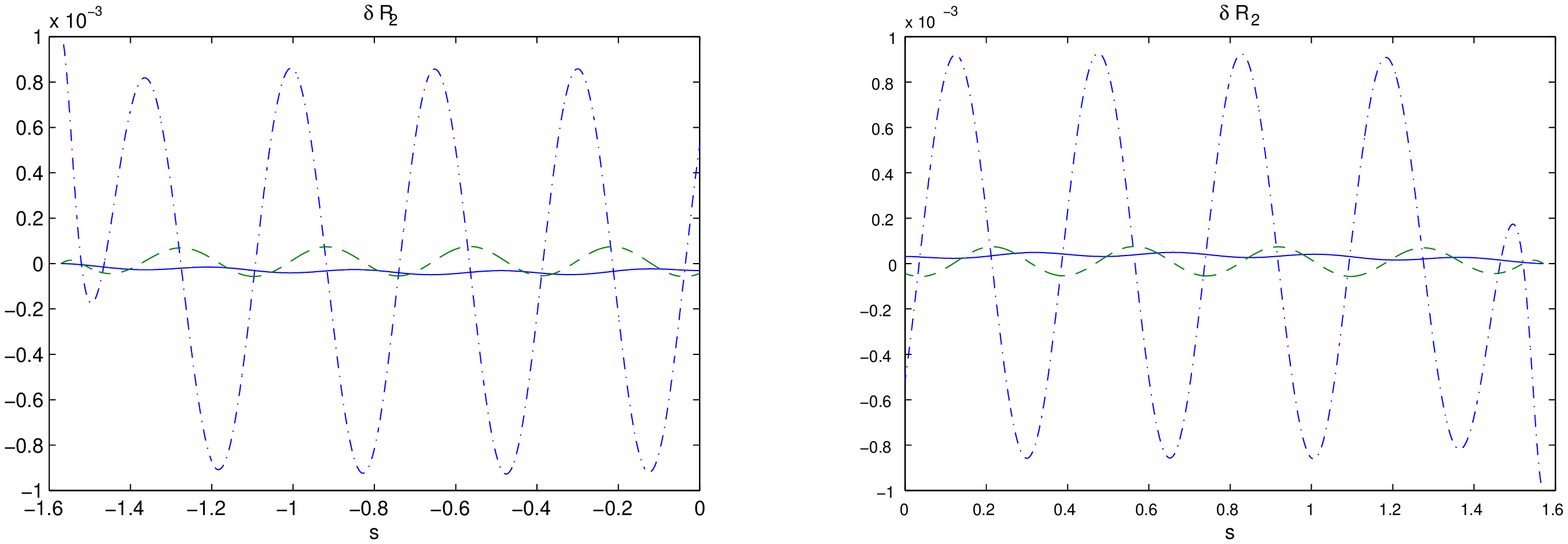}
 \caption{\footnotesize Variation of the  radius $R_2$ of the perturbed curve,
  with $r_0=1.5$ and $\beta=1$. The initial conditions for the  first graph are: a) 
   $\xi^{\mu}(-\frac{\pi}{2})=[0,0,0,0]$,
  $\dot{\xi}^{\mu}(-\frac{\pi}{2})=[10^{-3},0,0,0]$ (solid line),   b) 
  $\xi^{\mu}(-\frac{\pi}{2})=[0,0,0,0]$,
  $\dot{\xi}^{\mu}(-\frac{\pi}{2})=[0,10^{-3},0,0]$ (dot-dashed line), and c) 
  $\xi^{\mu}(-\frac{\pi}{2})=[0,0,0,0]$,
  $\dot{\xi}^{\mu}(-\frac{\pi}{2})=[0,0,10^{-3},0]$ (dashed line).
And initial conditions for the  second graph are:
 a)
  $\xi^{\mu}(\frac{\pi}{2})=[0,0,0,0]$,
  $\dot{\xi}^{\mu}(\frac{\pi}{2})=[-10^{-3},0,0,0]$ (solid line), b) the 
  $\xi^{\mu}(\frac{\pi}{2})=[0,0,0,0]$,
  $\dot{\xi}^{\mu}(\frac{\pi}{2})=[0,-10^{-3},0,0]$ (dot-dashed line), and 
  $\xi^{\mu}(\frac{\pi}{2})=[0,0,0,0]$,
  $\dot{\xi}^{\mu}(\frac{\pi}{2})=[0,0,-10^{-3},0]$ (dashed line).}
 \label{pert_figure1}
 \end{figure}

In Fig. \ref{pert_figure1} we show graphics that represent the
  variation of radius $R_2$ of perturbed curve, with $r_0=1.5
 (\eta\sim 10)$ and $\beta=1$.  For the  first graph the initial conditions
are: a) $\xi^{\mu}(-\frac{\pi}{2})=[0,0,0,0]$,
  $\dot{\xi}^{\mu}(-\frac{\pi}{2})=[10^{-3},0,0,0]$ (solid line),   b) 
  $\xi^{\mu}(-\frac{\pi}{2})=[0,0,0,0]$,
  $\dot{\xi}^{\mu}(-\frac{\pi}{2})=[0,10^{-3},0,0]$ (dot-dashed line), and c) 
  $\xi^{\mu}(-\frac{\pi}{2})=[0,0,0,0]$,
  $\dot{\xi}^{\mu}(-\frac{\pi}{2})=[0,0,10^{-3},0]$ (dashed line). And for the second 
are:
a)
  $\xi^{\mu}(\frac{\pi}{2})=[0,0,0,0]$,
  $\dot{\xi}^{\mu}(\frac{\pi}{2})=[-10^{-3},0,0,0]$ (solid line), b) the 
  $\xi^{\mu}(\frac{\pi}{2})=[0,0,0,0]$,
  $\dot{\xi}^{\mu}(\frac{\pi}{2})=[0,-10^{-3},0,0]$ (dot-dashed line), and 
  $\xi^{\mu}(\frac{\pi}{2})=[0,0,0,0]$,
  $\dot{\xi}^{\mu}(\frac{\pi}{2})=[0,0,-10^{-3},0]$ (dashed line).

\begin{figure}[!ht]
 \centering
 \includegraphics[scale=.5]{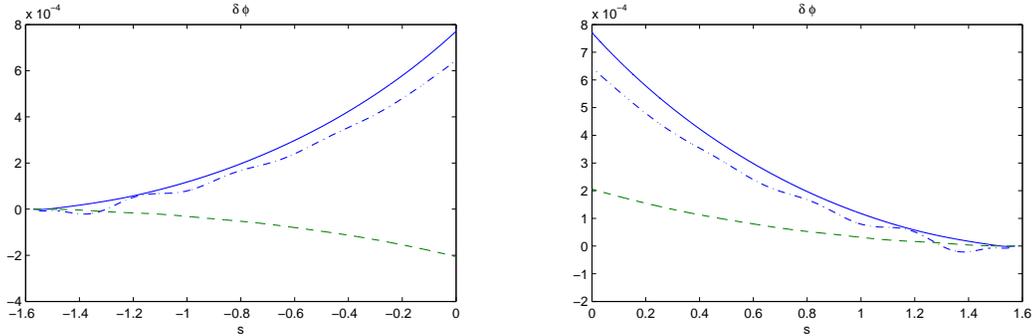}
 \caption{\footnotesize Variation of the angle $\phi$ for the perturbed
  curve with the same initial conditions of Fig.~\ref{pert_figure1}.}
\label{pert_figure2}
 \end{figure}

\begin{figure}[!ht]
 \centering
 \includegraphics[scale=.5]{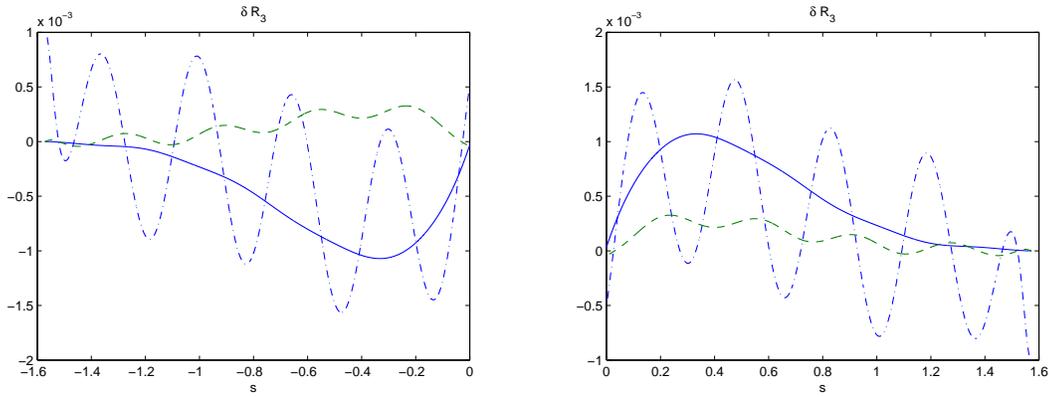}
 \caption{\footnotesize Variation of $R_3$  with same initial conditions  of Fig.~ \ref{pert_figure1}.}
\label{pert_figure3}
 \end{figure}

In Figs. \ref{pert_figure2} and \ref{pert_figure3}  we show 
graphics that represent $\delta \phi$  and $\delta R_3$, respecticely, 
computed  with the same initial conditions used for $\delta R_2$.

We consider a generic perturbation  $\delta u $  small  when 
$\frac{(\delta u)^2}{\delta u} \le 0.01$, i.e.,
$(\delta u)^2<<\delta u$. All the variations presented in the graphics 
satisfy by large this smallness conditions.
We have typical vibrational modes for $\delta R_2$ and $\delta  R_3$ and
mainly translational modes for $\delta \phi$. 
 We tested this smallness condition for a significant variety of
 initial conditions and values of $r_0$, we find  that this condition is 
always satisfied  even for curves that are near the bound given by
$r_0>\log(\sqrt{2}+1)$.

\section{ CTCs and matter}

The material source of the G\"odel metric is a pressure-free perfect
 fluid. In this section we  study the persistence and stability of
 the CTCs when we add matter with pressure or tension. 

Let us consider  the metric,
\begin{equation}
ds^2=(dt+\frac{\sqrt{2}h_1}{\beta
y}dx)^2-\frac{h_2^2}{(\beta y)^2}(dx^2+dy^2)-dz^2
\label{metricaPhgeral}
\end{equation}
that for $h_1=h_2=1$ reduces to G\"odel metric. This metric can be
considered as a deformation of the original G\"odel metric (See
Appendix A).

First, to understand the physical meaning of
the changes  introduced in (\ref{metricaPhgeral}) we compute the associated 
energy-momentum tensor from the Einstein field equations with the same cosmological  constant as in the  G\"odel metric ($\Lambda=-\beta^2/2$). 
We get,
 \begin{eqnarray}
\big(T^{\mu}_{\nu}\big) =\left[\begin{array}{cccc}
\frac{\beta^2(3h_1^2-2h_2^2+h_2^4)}{2h_2^4} & 
\frac{\sqrt{2}\beta h_1(2h_1^2-h_2^2)}{yh_2^4} &  0 &  0\\
0 & \frac{\beta^2(h_2^4-h_1^2)}{2h_2^4} & 0 & 0\\
0 & 0 & \frac{\beta^2(h_2^4-h_1^2)}{2h_2^4} & 0\\
0 & 0 & 0 & \frac{\beta^2(h_1^2-2h_2^2+h_2^4)}{2h_2^4}
\end{array}\right].
\label{TensorEMPgeral}
\end{eqnarray}

By solving the eigenvector equation,
\begin{equation}
T^{\mu}_{\nu}\xi^{\nu}=\lambda\xi^{\mu},
\end{equation}
we find the  eigenvalues: $\lambda_0=\dfrac{A\beta^2}{B}$,
$\lambda_1=\lambda_2=\dfrac{C\beta^2}{B}$ and
$\lambda_3=\dfrac{D\beta^2}{B}$, where $A=3h_1^2-2h_2^2+h_2^4$,
$B=2h_2^4$, $C=h_2^4-h_1^2$ and $D=h_1^2-2h_2^2+h_2^4$. The timelike
eigenvector $u^{\mu}= [1,0,0,0]$ is associated to $\lambda_0$, and the
spacelike eigenvectors
$X^{\mu}=[-\dfrac{\sqrt{2}h_1}{h_2},\dfrac{\beta y}{h_2},0,0], \;\;
Y^{\mu}=[0,0,\dfrac{\beta y}{h_2},0],$ and $Z^{\mu}=[0,0,0,1]$ are
associated to $\lambda_1,\;\; \lambda_2$, and $\lambda_3$,
respectively. We can write the energy-tensor in its canonical form as,
\begin{equation}
 T^{\mu \nu}=\lambda_0 u^\mu u^\nu+\lambda_1 (X^\mu X^\nu +Y^\mu
 Y^\nu)+\lambda_3(Z^\mu Z^\nu).
\end{equation}
In order to have realistic matter,  the eigenvalue $\lambda_0$
that represents energy density, and the eigenvalues
$\lambda_1=\lambda_2$ and $\lambda_3$ that describe pressures (or
tensions) are restricted.  $\lambda_0$ must  be non-negative (weak
energy condition) and  $\lambda_0-\lambda_1-\lambda_2-\lambda_3>0$ (strong
energy condition) \cite{hawking}.  The weak energy condition is
satisfied when $h_1^2>\frac{2h_2^2-h_2^4}{3}$. If, furthermore,  we
have $h_1^2>\frac{h_2^4}{2}$ the strong energy condition is also
satisfied.

Simple  modifications as the one presented in (\ref{metricaPhgeral})
of the G\"odel metric written in standard coordinates are not
associated with an energy-momentum tensor with a simple physical
interpretations nor  to geodesic equations that can  exactly be  solved. 
 This justify our use of Cartesian coordinates.

Returning to the analysis of the curves in this  new geometry, we have
that the geodesic equations are,
\begin{eqnarray}
&&\ddot{t}-\frac{2h_1^2}{yh_2^2}\dot{t}\dot{y}
-\frac{\sqrt{2}h_1}{\beta y^2h_2^2}(2h_1^2-h_2^2)\dot{x}\dot{y}=0\\ 
&&\ddot{x}+\frac{\beta\sqrt{2}h_1}{h_2^2}\dot{t}\dot{y}
+\frac{2\dot{x}\dot{y}}{yh_2^2}(h_1^2-h_2^2)=0\\
&&\ddot{y}-\frac{\beta\sqrt{2}h_1}{h_2^2}\dot{t}\dot{x}
-\frac{2h_1^2-h_2^2}{yh_2^2}(\dot{x})^2-\frac{1}{y}(\dot{y})^2=0\\
&&\ddot{z}=0.
\label{SistemaPhgeral}
\end{eqnarray}
By appropriated choice of the constants of integration, the first integrals
to these geodesic equations can be written as
~(\ref{first_integral_t})-(\ref{first_integral_z}) replacing $\dot{t}$
by
\begin{equation}
\dot{t} = \frac{1}{\sqrt{2}Ch_1}[2yh_1^2-y_0(2h_1^2-h_2^2)].
\label{1integraisPhgeral}
\end{equation}
Therefore
\begin{equation}
t=\frac{\sqrt{2}}{\beta}[2\arctan(\sqrt{N}\tan\
\sigma)h_1- \frac{(2h_1^2-h_2^2)}{h_1}\frac{(N+1)}{2\sqrt{N}}\sigma]+x_0^0.
\label{solucaoGeodesicaG3Phgeral}
\end{equation}
We have   timelike geodesics for 
\begin{equation}
y_0=\cosh\  2r_0<\frac{\sqrt{2}h_1}{\sqrt{2h_1^2-h_2^2}},
\end{equation}
that are closed whenever
\begin{equation}
y_0=\cosh\  2r_0=\frac{2h_1^2}{2h_1^2-h_2^2}.
\end{equation}
Hence, as before, we have no closed timelike geodesics in the spacetime whose 
metric is given by~(\ref{metricaPhgeral}).
In this metric  we can find equations for CTCs  in the same way as
before. For $z=0$, we have

\begin{equation}
t = \frac{2\sqrt{2}}{\beta}\left[
\arctan\big(\sqrt{N}\tan\sigma\big)h_1-
\frac{\sqrt{2}y_0h_1-\sqrt{C^2+h_2^2\eta^2}}{\sqrt{2}\sqrt{y_0^2-\eta^2}}
\sigma\right]+t_0.
\label{solucaoCTCG3hgeral}
\end{equation}
The condition for closed curves is
\begin{equation}
C^2=2h_1^2(y_0-\sqrt{y_0^2-\eta^2})^2-h_2^2\eta^2.
\label{conditionCTCgeneral}
\end{equation}
The components of force are,
\begin{eqnarray}
&&F^0_p =  -(x-x_0)\frac{\lambda(h_1,h_2) \sqrt{2}}{\beta},\\
&&F^1_p =  (x-x_0)\  y\  \lambda(h_1,h_2), \\
&&F^2_p = -y\  \lambda(h_1,h_2)(y_0-y), \\
&&F^3_p =0,
\label{forcaG3hgeral}
\end{eqnarray}
where
\begin{equation}
\lambda(h_1,h_2)=\dfrac{\beta^2(\sqrt{2}
\sqrt{C^2+h_2^2\eta^2}h_1-y_0h_1^2)}{C^2h_2^2}.
\label{lambdah1h2}
\end{equation}

We have CTCs when  $C^2> 0$. From (\ref{conditionCTCgeneral}) we find the  relation between
$h_1$ and $h_2$ to have CTCs, 
\begin{equation}
h_1^2 > \frac{\eta^2 h_2^2}{2(y_0-1)^2}.
\label{h1andh2}
\end{equation}
 
From  the equations for the  CTCs it is possible to analyze its
stability under linear perturbation. As before, we write the
coordinates of the perturbed curve as
$\tilde{X}^{\mu}=X^{\mu}+\xi^{\mu} $ and find a system 
like (\ref{systemPerturbation}), where $(a_{ij}), \; (b_{ij})$  are  now  given by
\begin{eqnarray}
(a_{ij}) = \frac{1}{Ch_2^2}\left[ \begin{array}{ccc}
  \frac{4 \beta S \ee h_ 1^2}{D}                  &
  2\sqrt{2}\,S\,h_1N_1                 & 
  \frac{2\sqrt{2}h_1(2h_1-N_1+(y_0N_1-2h_1)De^{2r_0})}{2}\\[5mm]
 -\frac{2\sqrt{2}\,\beta^2 S e^{-4r_0}h_1}{D}      & 
 -\frac{4\,\beta\,S e^{-2r_0}N_2}{D^2}  & 
 \frac{-2\beta(\ee (N_2+h_1)-D(h_1+N_2y_0))}{D}\\[5mm]
  \frac{\sqrt{2}\,\beta^2 \ee h_1(y_0D-e^{-2r_0})}{D^2}        & 
  \frac{2\beta(\ee (h_1-N_2)-D(h_1^2-N_1y_0)}{D} &
  \frac{2 \beta S \ee}{D}
\end{array}\right],
\end{eqnarray}

\begin{equation}
(b_{ij}) = \left[ \begin{array}{cc}
\displaystyle  - \frac{\lambda\sqrt{2}}{\beta}                 &
\displaystyle  - \frac{\ee(N_1-h_1)+D(h_1-N_1y0)}{Dh_2^2C^2}4\sqrt{2}\beta S                   \\[5mm]
\displaystyle  \frac{\ee\;\lambda}{D}                        & 
\displaystyle  \frac{\ee\;S}{D}\left[ \frac{4\beta^2N_2}{h_2^2C^2}
(\frac{\ee}{D}-y0)+2\lambda\right]              \\[5mm]
\displaystyle  0                                               & 
\displaystyle  \lambda \left[ \frac{2\ee}{\dd}-y_0 \right]-\frac{
\beta^2}{h_2^2C^2}\left[N_1(\frac{\ee}{D}-y0)^2+\frac{4h_2^2\;S^2
 e^{-4r_0}}{D^2} \right]
\end{array}\right],
\end{equation}
 $N_1=2h_1^2-h_2^2$, $N_2=h_2^2-h_1^2$, and  $\lambda$ is like in
(\ref{lambdah1h2}). As before, $\ddot{\xi}^3=0$ and we do $\xi^3=0$.

\begin{figure}[!ht]
 \centering \includegraphics[scale=.5]{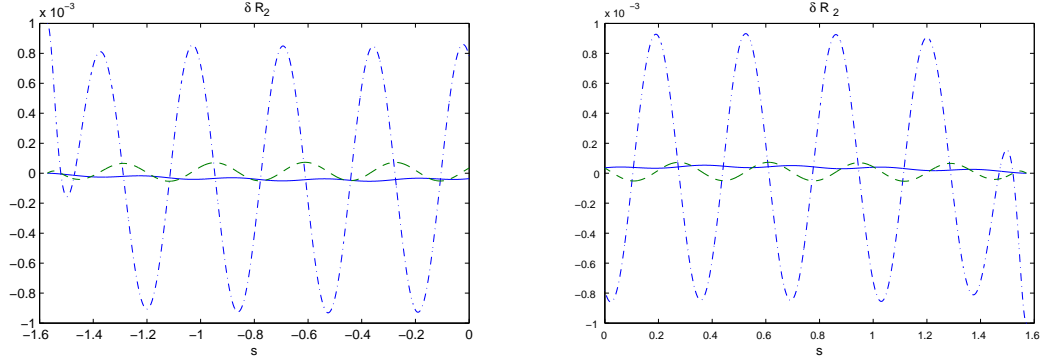}
 \caption{\footnotesize Similar curves as in the
   Fig.\ref{pert_figure1} with the same initial conditions. But, now
we have  ordinary matter
   with parameters $h_1=\sqrt{1.1}$ and $h_2=\sqrt{1.1}$.}
 \label{pert_figure4}
 \end{figure}

\begin{figure}[!ht]
 \centering
 \includegraphics[scale=.5]{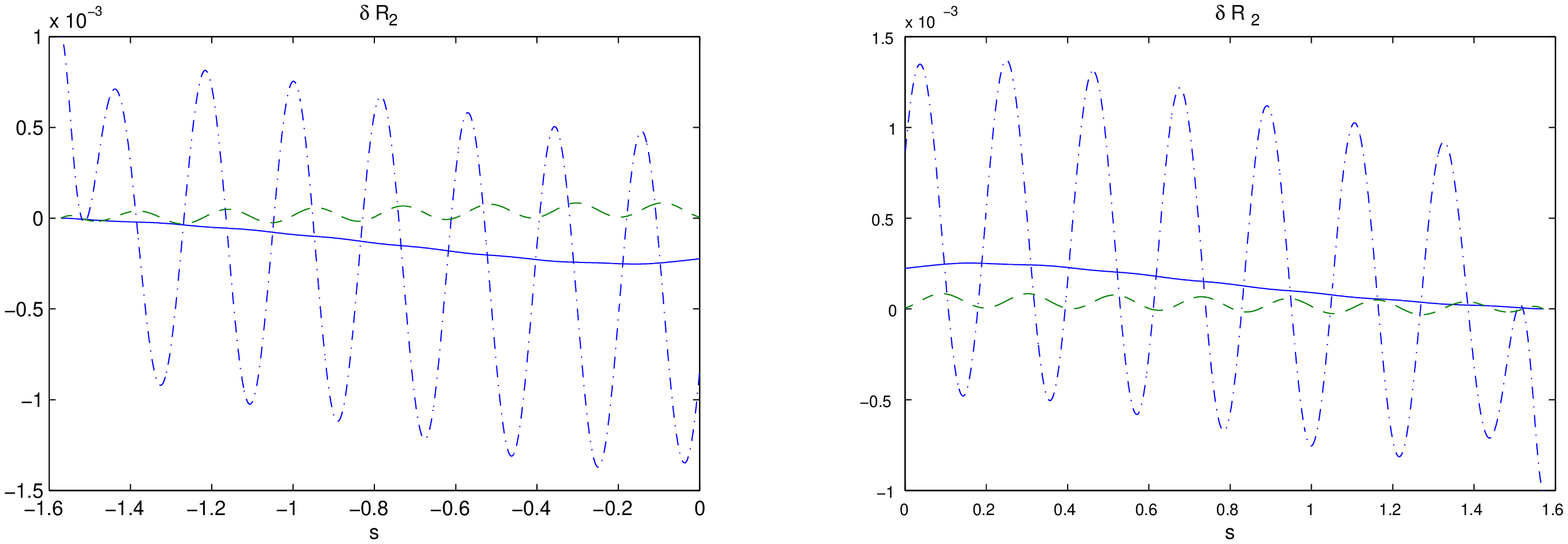}
 \caption{\footnotesize Similar curves as in the
   Fig. \ref{pert_figure1}  with the same initial conditions. But, now
we have  exotic matter
   with paramerter  $h_1=\sqrt{2.7}$ and $h_2=\sqrt{2.4}$.}
 \label{pert_figure5}
 \end{figure}
 
 The analysis of the linear stability of CTCs in this new universe is
 made in the same way as in the  G\"odel's case. Several initial
 conditions are tested and  the results for usual as well as for exotic matter
 are very similar to those obtained before.
 We tested the stability of CTCs in various scenarios.
In Fig. ~\ref{pert_figure4}  we plot the function $\delta R_2$ for the same  initial
conditions used in Fig. \ref{pert_figure1} and $h_1=\sqrt{1.1},\;h_2=\sqrt{1.1}$.  We see a similar behavior as in the pure G\"odel case. In Fig. \ref{pert_figure5} 
we show  $\delta R_2$  for the same initial conditions for exotic
matter ($\lambda_0-\lambda_1-\lambda_2-\lambda_3<0$)  with $h_1=\sqrt{2.7},\;h_2=\sqrt{2.4}$. We see that  the linear  stability is kept.

For some values of $h_1$ and $h_2$ the condition $C^2>0$ [see, Eq.
(\ref{conditionCTCgeneral})] is not true.  Therefore it is possible to have
spacetimes  with a metric like (\ref{metricaPhgeral})  with no  CTCs 
for $r_0> \log(1+\sqrt{2})$ and any kind of matter  (ordinary or exotic).

\begin{figure}[!ht]
 \centering
 \includegraphics[scale=.6]{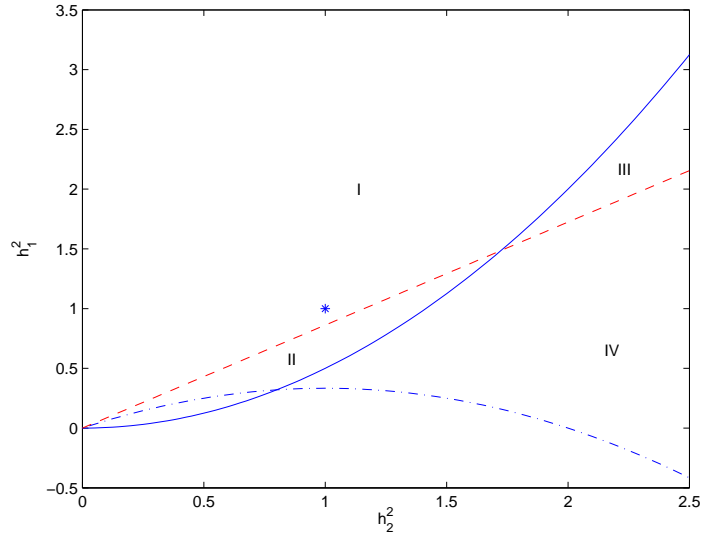}
 \caption{\footnotesize 
 For values of $(h_2^2,h_1^2)\in I$ we have   spacetimes  with  CTCs and
ordinary matter. When   $(h_2^2,h_1^2) \in II$   the spacetime contains
ordinary matter, but not  CTCs. If  $(h_2^2,h_1^2) \in III$  the
spacetime contains CTCs and exotic matter.  And  for $(h_2^2,h_1^2) \in IV$ the spacetime contains exotic matter, but not CTCs. The isolated
point represent the G\"odel universe $(h_1,h_2)=(1,1)$. }
 \label{graphicsH1H2}
 \end{figure}

 In Fig. \ref{graphicsH1H2} three lines  divide the
$(h_2^2,h_1^2)$-plane in four regions of  interest that  are
marked with $I$, $II$, $III$ and $IV.$ 
 In the region above the dot-dashed line  [$h_1^2=(2h_2^2-h_2^4)/3$]
  we have  matter with positive energy density.  
Above the solid line [$h_1^2=(h_2^4)/2$] the strong energy
 condition is fulfilled.  
Under the solid line   we have
  exotic matter. The dashed line (straight line)
separates  the values of $h_1$ and $h_2$ for  spacetimes with CTCs
(above) to the ones that do not have CTCs (bellow).
This line is    $h_1^2=0.862030830483155 h_2^2$ that  is obtained replacing $r_0=1$ 
  in  (\ref{h1andh2}).  

Therefore, when  $(h_2^2,h_1^2) \in I$ the  spacetime contains CTCs and
ordinary matter. For   $(h_2^2,h_1^2) \in II$   the spacetime contains
ordinary matter, but not  CTCs. If  $(h_2^2,h_1^2) \in III$  the
spacetime contains CTCs and exotic matter.  And  when $(h_2^2,h_1^2) \in
IV$ the spacetime contains exotic matter, but not CTCs. The isolated
point represent the G\"odel universe $(h_1,h_2)=(1,1)$. As we can see
in this figure we have an open neighborhood  of the point $(1,1)$ (G\"odel spacetime)
where the matter is ordinary, the CTCs are present and  are  stable as  depicted  in Fig. \ref{graphicsH1H2}.

Loosely speaking, we have structural stability of a vector field when
the equations that it satisfies are slightly changed we also have  a 
small change in the trajectories represented by the vector field 
 (for  a simple introduction to this subject see reference
\cite{glendinning}). We see that the addition of matter with constants
$h_1\sim 1$ and $h_2 \sim 1$ changes ``slightly'' the original
geodesic equations for the G\"odel metric. We still have closed curved that are similar
to the original G\"odel CTCs. Therefore we can say that the G\"odel
CTCs are structural stable under the inclusion of the special matter
represented by the energy-momentum tensor (\ref{TensorEMPgeral}).

\section{Conclusions}

In this work we verify that the closed timelike curves in
G\"odel spacetime are stable under linear perturbations.
We also show that the G\"odel spacetime has a stable structure under a
special class of deformations.  We found explicit  CTCs in 
the deformed spacetime and proved that closed timelike
geodesics do  not exist. The
energy-momentum tensor of the deformed spacetime was studied in some detail,
specially we examined  the conditions to have  exotic and usual  matter. We studied the stability of the new CTCs under
linear perturbation and found that these curves are also
stable. We tested these curves in a spacetime with exotic matter
and find the same properties of stability as in the case of ordinary matter.
We also find a kind of structural stability of the CTCs.

\acknowledgments
V.M.R.  thanks Departamento de Matem\'atica-UFV for giving
the conditions to finish this work which was partially supported  by
PICDT-UFV/CAPES. PSL thanks the partial financial support of FAPESP and
CNPq.

\vspace{0.5cm}

\begin{center}
{\bf Appendix A}
\end{center}
\vspace{0.5cm}

 G\"odel in his seminal article~\cite{goedel} mentions that his 
 metric has five isometries~\cite{goedel};
 Kundt~\cite{kundt} shows explicitly four out of the five
Killing vectors of the above mentioned metric in Cartesian coordinates, 
\begin{equation}
\zeta_{(0)}=\frac{\partial}{\partial
t},\;\zeta_{(1)}=\frac{\partial}{\partial
x},\;\zeta_{(3)}=\frac{\partial}{\partial z},
\end{equation}
\begin{equation}
\zeta_{(2)}=x\frac{\partial}{\partial x}+y\frac{\partial}{\partial y}.
\end{equation}
 The fifth Killing vector is not so  trivial we find,
\begin{equation}
\zeta_{(4)}=\frac{\sqrt{2}}{\beta}(y-1)\frac{\partial}{\partial
t}+\frac{1}{2}(1-y^2+x^2)\frac{\partial}{\partial
x}+xy\frac{\partial}{\partial y}.
\end{equation}
For a discussion of the topology of G\"odel metric and the five
Killing vectors in a  different system of coordinates,
see~\cite{soares}. 

The five G\"odel Killing vectors plus the discrete  symmetry of reflection on the 
$z=0$ plane  ( $z\rightarrow -z$)  give
us a family of metrics, that  we will named the  G\"odel family,
\begin{equation}
ds^2=k_1dt^2+\frac{2k_2}{y}dtdx+\frac{k_3}{y^2}dx^2+\frac{k_4}{y^2}dy^2+k_5dz^3;
\end{equation}
where $k_i, i=1,\dots,4$ satisfy the following relations:
\begin{equation}
k_2-\frac{\sqrt{2}}{\beta}k_1=0,\;\frac{\sqrt{2}}{\beta}k_2-k_3+k_4=0.
\end{equation}

The corresponding constants that appear in the metric
 (\ref{metricaPhgeral}) do not satisfy the above relation. In fact,
 only four of the above mentioned five Killings vectors ($\zeta_{(0)},
 ... ,\zeta_{(3)})$ are symmetries of (\ref{metricaPhgeral}),
 therefore this last metric can be considered as a deformation of the
 G\"odel metric.



\begin{thebibliography}{99}

\bibitem{goedel}K.G\"odel,  {\it Rev.Mod.Phys.}, {\bf  21}, 447(1949)

\bibitem{hawking}S.W.Hawking  and G.F.R.Ellis,  {\it The Large
Structure of Space-Time} (Cambridge University Press, Cambridge,
1973)

\bibitem{pfarr}J.Pfarr, {\it Gen.Rel.Grav.}, {\bf 13}, 1073(1981)

\bibitem{gleiser}R.J.Gleiser,  M.G\"urses,  A.Karasu, O.Sarioglu,
 {\it Class.QuantumGrav.}, {\bf 23}, 2653(2006)

\bibitem{calvao} M.O.Calv\~ao,  I.D.Soares,  J.Tiommo, {\it
Gen.Rel.Grav.}, {\bf 22}, 683(1990)

\bibitem{gurses} M.G\"urses,  A.Karasu,  O.Sarioglu, {\it
Class.Quantum.Grav.}, {\bf 22}, 1527(2005)

\bibitem{kundt}W.Kundt,  {\it Z.Phys.}, {\bf 145}, 611(1956)

\bibitem{chandra} S.Chandrasekhar and J.P.Wright, {\it
Proc.Natl.Acad.Sci.U.S.A}, {\bf 47}, 341(1961)

\bibitem{klepac}P.Klepac and J.Horrsky,  {\it
Class.Quantum.Grav.}, {\bf 17}, 2547(2000)

\bibitem{gersl}J.Gersl, P.Klepac, J.Horsky,  {\it
Gen.Rel.Grav.}, {\bf 36}, 1399(2004)

\bibitem{ivanov}B.V.Ivanov,  {\it Class.Quantum.Grav.}, {\bf
19}, 5135(2002)

\bibitem{ozdemir} N.Ozdemir, {\it Int.J.Mod.Phys.A}, {\bf
20}, 2821(2005)
\bibitem{gott} J.R. Gott, III, {\it Phys. Rev. Lett. } 66, 1126 (1991) 

\bibitem{barrow}J.D.Barrow, and  C.G.Tsagas,
  {\it Class.QuantumGrav.}, {\bf 21}, 1773(2004)

\bibitem{novello}M.Novello,  I.D.Soares,  J.Tiommo, {\it
Phys.Rev.D}, {\bf 27}, 1399(1983)

\bibitem{hawking2} S.W.Hawking, {\it Phys.Rev.D}, {\bf 46}, 603(1992)

\bibitem{bonnor2} W.B.Bonnor, {\it Class.Quantum.Grav.}, {\bf
20}, 3087(2003)

\bibitem{bonnor} W.B.Bonnor, {\it Class.QuantumGrav.}, {\bf
19}, 5951(2002)

\bibitem{shirokov} M.F.Shirokov, {\it Gen.Rel.Grav.}, {\bf 2}, 131(1971)

\bibitem{soares} I.D.Soares, {\it Phys.Rev.D}, {\bf 23}, 272(1981)

\bibitem{glendinning} P. Glendinning, {\it Stability, instability and
  chaos: an introduction to the theory of nonlinear differential
  equations} (Cambridge University Press,  1996)


\end{thebibliography}
\end{document}